%% file: main.tex
\documentclass[letterpaper,twocolumn,10pt]{article}
\usepackage{usenix,epsfig,endnotes}

\usepackage{cite, url}
\usepackage{times}
\usepackage{amsmath}
\usepackage{graphicx}
\usepackage{graphics}
\usepackage{epsfig}
\usepackage{latexsym}
\usepackage{amsfonts}
\usepackage{amssymb}
\usepackage{paralist}
\usepackage{xspace}
\usepackage{mathrsfs}
\usepackage{amssymb}
\usepackage{color}
\usepackage{algorithm}
\usepackage{algorithmic}
\usepackage{listings}
\usepackage{multirow}
\usepackage{pifont}
\usepackage{booktabs}
\usepackage{bbding}

\usepackage{epstopdf}
\usepackage{subfigure}
\newcommand{\cmark}{\ding{51}}%
\newcommand{\xmark}{\ding{55}}%

\newcommand*{\affaddr}[1]{#1} % No op here. Customize it for different styles.
\newcommand*{\affmark}[1][*]{\textsuperscript{*}}

\begin{document}

% What is going to be the name
\title{RDMAvisor: Toward Deploying Scalable and Simple RDMA as a Service in Datacenters \vspace{-.5cm}}
\def\name{RaaS}
\def\fullname{RDAM as a Service}
\def\sysname{{\tt \small RDMAvisor}}

%% \vspace*{-0.1in}
% \numberofauthors{1}
\author{\rm Zhi Wang, Xiaoliang Wang, Cam-Tu Nguyen, Zhuzhong Qian, Baoliu Ye, Sanglu Lu \\
\affaddr{ Nanjing University}     \quad
\affaddr{ Huawei} \\
\affaddr{ submit to [USENIX ATC '17] Paper \#9} 
}

\maketitle

%
% import the customized commands
%
\input{custom}
%
% section by section tex files
%
\input{abstract}
\input{intro}

% \input{background}
\input{design}
% \input{impl}
\input{eval}

% \input{related}
\input{conclude}
% \small
% \clearpage
\bibliographystyle{abbrv}
\bibliography{reference}

\end{document}

%% file: custom.tex
\newcommand{\knote}[1]
{{$\langle${Kun: \textbf{#1}}$\rangle$}}
\newcommand{\kfnote}[1] {\footnote{\textcolor{red}{Kun: \it{#1}}}}
\newcommand{\todo}[1] {\textcolor{red}{To do: \it{#1}}}
\newcommand{\remove}[1] {{\textcolor{red}{\sout{#1}}}}
\newcommand{\add}[1] {{\textcolor{red}{\underline{#1}}}}
\newcommand{\egg}[1] {}
\newcommand{\separate}[1] {\textbf{\center ======  #1  ====== }}
\newcommand{\smalltitle}[1] {\vspace{6pt} \noindent \textbf{#1}}

\def\ie{\textit{i.e.}\xspace}
\def\etal{\textit{et al.}\xspace}
\def\etc{\textit{etc.}\xspace}
\def\eg{\textit{e.g.}\xspace}
\def\st{\xspace\textbf{s.t.}\xspace}
\def\whp{{\emph{w.h.p}}}
\def\cname{FICA \xspace}
\def\802{IEEE 802.11\xspace}
\def\arrow{$\rightarrow$}
\def\mrts{\textit{M-RTS}\xspace}
\def\mrtss{\textit{M-RTSes}\xspace}
\def\mcts{\textit{M-CTS}\xspace}
\def\mctss{\textit{M-CTSes}\xspace}
\def\approx{$\sim$}

\def\receivers{\textbf Q}
\def\capacity{{ \textbf{c}}}
\def\upperBound{{\mathcal D}}
\def\schedule{{\mathcal S}}
\def\throughput{{\cal T}}
\def\period{{\textbf Z}}

\newtheorem{theorem}{Theorem}
\newtheorem{axiom}[theorem]{Axiom}
\newtheorem{corollary}[theorem]{Corollary}
\newtheorem{definition}{Definition}
\newtheorem{lemma}[theorem]{Lemma}
\newtheorem{remark}[theorem]{Remark}

\newcommand{\prob}[1]{{\textbf{Pr}\left(#1\right)}}
\newcommand{\mcell}[2]{ \parbox[h]{#1}{ \vspace{0.5mm} #2 \vspace{0.5mm}}}
\newcommand{\ns}[1]{\textit{ns#1}}

\lstdefinestyle{numbers} {numbers=left, stepnumber=1,
numberstyle=\tiny, xleftmargin=10pt, numbersep=5pt}
\lstset{language=C++}

%% file: abstract.tex
%!TEX root=main.tex
\begin{abstract}
%%
%%  here is the abstract
%%
RDMA is increasingly adopted by cloud computing platforms to provide low CPU overhead, low latency, high throughput network services. On the other hand, however, it is still challenging for developers to realize fast deployment of RDMA-aware applications in the datacenter, since the performance is highly related to many low-level details of RDMA operations. To address this problem, we present a simple and scalable RDMA as Service (RaaS) to mitigate the impact of RDMA operational details. RaaS provides careful message buffer management to improve CPU/memory utilization and improve the scalability of RDMA operations. These optimized designs lead to simple and flexible programming model for common and knowledgeable users. We have implemented a prototype of RaaS, named {\sysname}, and evaluated its performance on a cluster with a large number of connections. Our experiment results demonstrate that {\sysname} achieves high throughput for thousand of connections and maintains low CPU and memory overhead through adaptive RDMA transport selection. 
\end{abstract}

%% file: intro.tex
%!TEX root=main.tex
\section{Introduction}
\label{sec:intro}

\subsection{Background and Motivation}
Remote Direct Memory Access (RDMA) technique provides the messaging service that directly access the virtual memory on remote machines. Since data can be copied by the network interface cards (NICs), RDMA provides minimal operating system involvement and achieves low latency data transportation through stack bypass and copy avoidance. It has been widely used by the high performance computing (HPC) community and closely coupled with the InfiniBand (IB) network. 

% and releases precious CPUs resource for the computation services

Recently, due to the decreasing price of RDMA hardware and the compatible design of RDMA over Ethernet (RoCE) standard, the distributed computing platforms have been testing and deploying RDMA in order to alleviate communication bottleneck of existing TCP/IP-based environment \cite{w15gram, z15congestion}. To make RDMA network scalable to support hundreds of thousands of nodes in the datacenter, the IP routable RoCE (RoCEv2) \cite{rocev2} protocol was also defined and quickly evaluated in the Microsoft datacenter \cite{z15congestion, g16rdma}. By leveraging the advanced techniques of priority-based flow control (PFC) \cite{pfc} and congestion notification (QCN) \cite{z15congestion} it has indicated the potential RDMA by replacing TCP for intra-datacenter communications \cite{g16rdma}. 

% achieve low latency, low CPU overhead, and high throughput ranging from 10Gb/s up to 100Gb/s 

RDMA defines asynchronous network programming interfaces, RDMA verbs, for submitting work requests to the channel adapter and returning completion status. Three transport types are provided: Reliable Connection (RC), Unreliable Connection (UC), and Unreliable Datagram (UD). Reliable transport guarantees lossless transfer and ensures in-order delivery of messages by using acknowledge. The unreliable transport providing no sequence guarantee consumes less bandwidth and delay due to no ACK/NACK packet. In the RDMA verbs, both channel semantics and memory semantics are provided to users. The channel semantics are two-sided operations, sometimes called \emph{Send/Receive}, which is the communication style used in a classic I/O channel. The Memory verbs are one-sided operations, embedded in the \emph{Read and Write} operations, which allows the initiator of the transfer to specify the source or destination location without the involvement of the other endpoint CPUs. Table \ref{tab:opcode} shows the operations available in each transport mode. 

\begin{table}
	\centering
\begin{tabular}{lcccc}
     & SEND/RECV  & WRITE    & READ     & Max Message\\
\hline
RC   & \cmark     & \cmark   & \cmark   & 1GB \\ 
UC   & \cmark     & \cmark   & \xmark   & 1GB \\
UD   & \cmark     & \xmark   & \xmark   & MTU 
\end{tabular}
  \caption{Operations and maximum message size supported by each transport type. RC and UC support message size up to 1GB. The message is divided into MTU-sized frames during transmission on Ethernet. UD supports the maximum size of a message up to MTU size.}
	\label{tab:opcode}
\end{table} 

\subsection{Challenges and Our Solution}
Despite rich transport modes and operations have been provided, it is still a challenging task to achieve advance capabilities for applications deploying in the RDMA-capable datacenters. With regard to the specific demands of different applications and the shared environment of datacenter, applying the naive RDMA may not be a wise move \cite{n15latency, b16end, k16fasst, k16design}. The best-fit RDMA-aware system design for a specific application requires careful parameters turning which highly relies on the low-level knowledge associated with the NIC architecture, operation details, etc. \cite{w14chint, d14farm, c16fast, k16design, k16fasst}. In general, the developer meets the following problems when migrating applications to RDMA-aware systems. 

\textbf{The first challenging is the operational availability.} Currently, the available design of RDMA network functionality is tightly coupled with the targeting system. The basic units of RDMA network functionality, e.g., RDMA enabled buffers management, data structures and polling thread, usually work as modules embedded in the system. The coupled design makes it hard to reuse the code in other applications. In addition, since the low-level details are important for RDMA system design \cite{k16design}, the application performance is highly determined by the choice of RDMA operations. Inappropriate configurations may cause even worse performance than the TCP/IP stack. Unfortunately there is no one-size-fits-all approach. The optimal setting of a given system can be the poison to other ones. It is not easy to train the system designers to master low-level details of RDMA technologies and understand the guidelines \cite{guidelines, k16design} for various conditions in a short time, which hind the fast development of RDMA-based system in datacenters. Thus, it is necessary to ensure that the application is not sensitive to the RDMA network operations by introducing a shim layer to mitigate the influence of low-level details. To address this challenge, the flexible RDMA functions are abstracted by Socket-like network interfaces. All the conditional sections in the program are described with parameters. For common users, the low-level details are not exposed and the performance is guaranteed by adaptive RDMA primitive selection in the shim layer. For specific requirements their demands can be realized through customized setting through macro arguments. 

% Mellanox's Dynamically Connected Transport (DCT) \cite{producebrief} 
\textbf{The second challenging is the scalability.} Scalability is a practical concern for modern datacenters where a large number of connections are generated in one machine and also requires for incremental expansion. However, due to the limited cache space, currently ConnectX-3 NIC can only supports around 400 Queue Pairs (QPs). Thus, the connected transport which provides exclusive access to QPs is not scalable. QPs sharing is a common solution to this problem \cite{producebrief}. For small messages, key-value store \cite{k14keyvalue} and RPC \cite{k16fasst} leveraged the two-side verbs, which allow each CPU core to create one datagram QP that can communicate with all remote cores. However, Send/Recv operations require the involvement of CPUs at both the sender and receiver. For large items, QP sharing for one-side verbs was implemented in \cite{d14farm}. But it can reduce CPU efficiency because threads contend for locks \cite{k16fasst}. To address this challenge, a more efficient lock-free design is provided in our design for one-side verbs to achieve high scalability but low CPU overhead.

\textbf{The third challenging is the low utilization of resource} when considering the shared datacenter environment. The one-side Recv operation needs to posts receive work requests (WRs) beforehand to handle incoming messages. Holding a lot of WQEs in receive queues (RQs) may result in inefficient memory or wire usage. Besides, data sink consumer may be unaware that the RQ is starving. The corresponding polling thread leads to the waste of CPU resource \cite{d14farm}. The SRQ model partially solve this problem which posts receive WRs to a queue that is shared by a set of connections. However, under the naive RDMA consumer queuing model, each application maintains and manages its own RDMA queues. Our key observation is that the resource such as SRQs can be shared among multiple applications that are running on the same physical machine. To this end, we provide careful memory management among transport connections of consumers that achieves an high utilization of memory usage, low CPUs overhead for the low latency, high throughput RDMA network service. 

\subsection{Contribution and Results}
The contribution of this paper is two fold. First, we present a simple and scalable \emph{RDMA as a Service (RaaS)} for fast deploying RDMA-capable services in the datacenter. RaaS works as a daemon process for all applications running on a physical machine. The interface semantics are fairly straightforward and friendly to developers. The user only need to decide to use connection or datagram transport service, because RaaS has mitigated the impact of low-level details by improving the scalability of one-sided operations, the resource utilization of two-sided operations and providing adaptive RDMA transport based on the usage of CPUs and memory at both end-hosts. Second, we have implemented a RaaS prototype {\sysname}, and evaluated the performance in a small cluster running a large number of connections. {\sysname} shows more efficient usage of memory and CPUs, and achieves high throughput for thousand of connections.

%% file: design.tex
%!TEX root=main.tex

\section{RaaS} \label{sec:design}

\subsection{Communication Primitives}
The foundation of RDMA operation is the ability of a consumer to queue up a set of instructions that hardware execute. This facility is referred to as a Work Queue. Work queues are always created in pairs, called a Queue Pair (QP), one for send operations and one for receive operations. The consumer submits a work request (WR), which causes an instruction called Work Queue Element (WQE) to be placed in the work queue. RDMA NIC executes WQEs in the order without involving the kernel. When finishing a request, a Completion Queue Element (CQE) is placed in the completion queue (CQ). 

Unreliable Datagram (UD) has good scalability, since the connectless service allows use one QP to communicate with multiple other QPs. UD operates with less state maintained at each end-host is suitable for latency-sensitive applications with small messages, such as key-value store \cite{k14keyvalue} or RPC service \cite{k16fasst}. UD Send/Receive operations require involvement of CPU at both sender and receiver. Besides, since the maximum message length is constrained to fit in a single packet (MTU), UD messages are conveyed by multiple packets, which is not suitable for throughput-sensitive applications. 

\begin{figure} \centering    
	\subfigure { \label{fig:optsa} 
		\includegraphics[width=0.46\columnwidth]{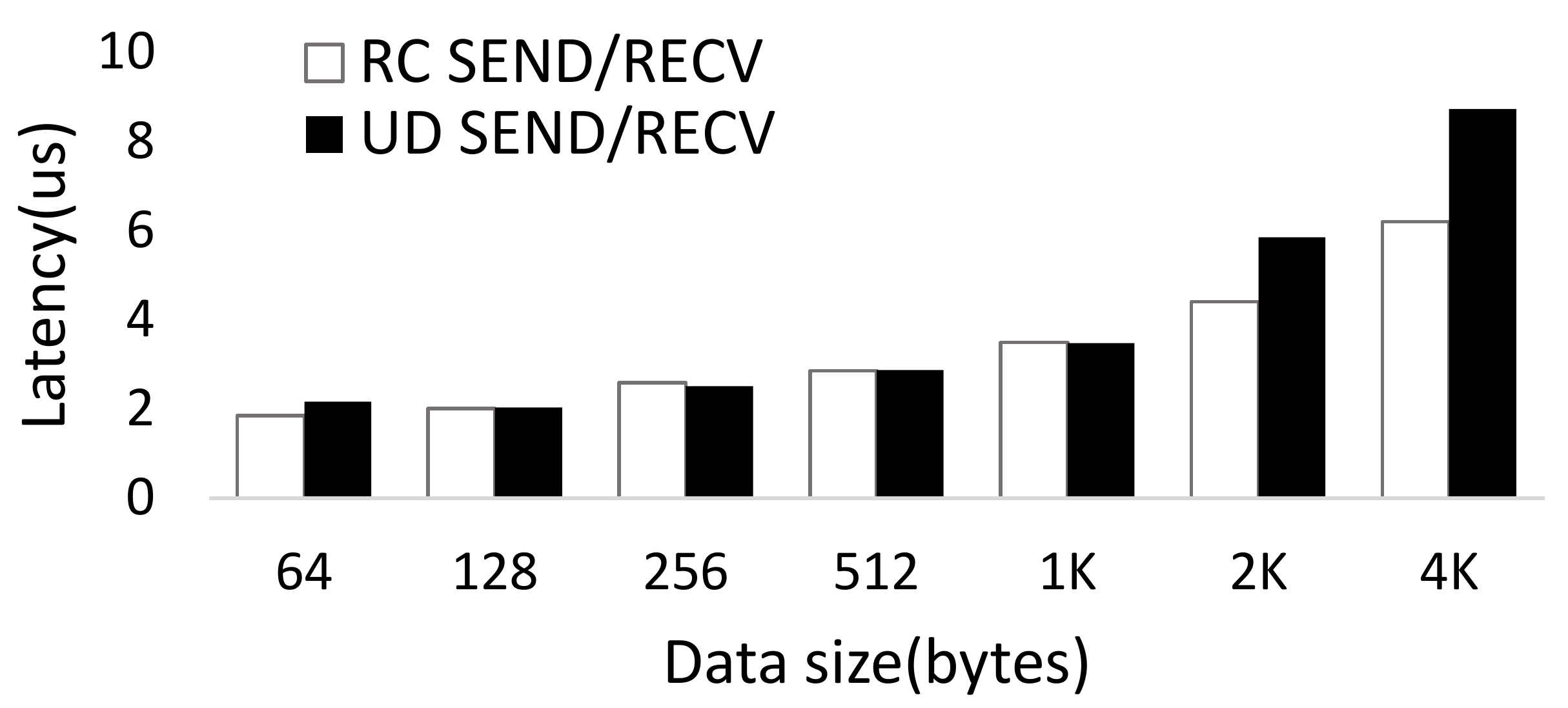}  
	}     
	\subfigure { \label{fig:optsb}     
		\includegraphics[width=0.46\columnwidth]{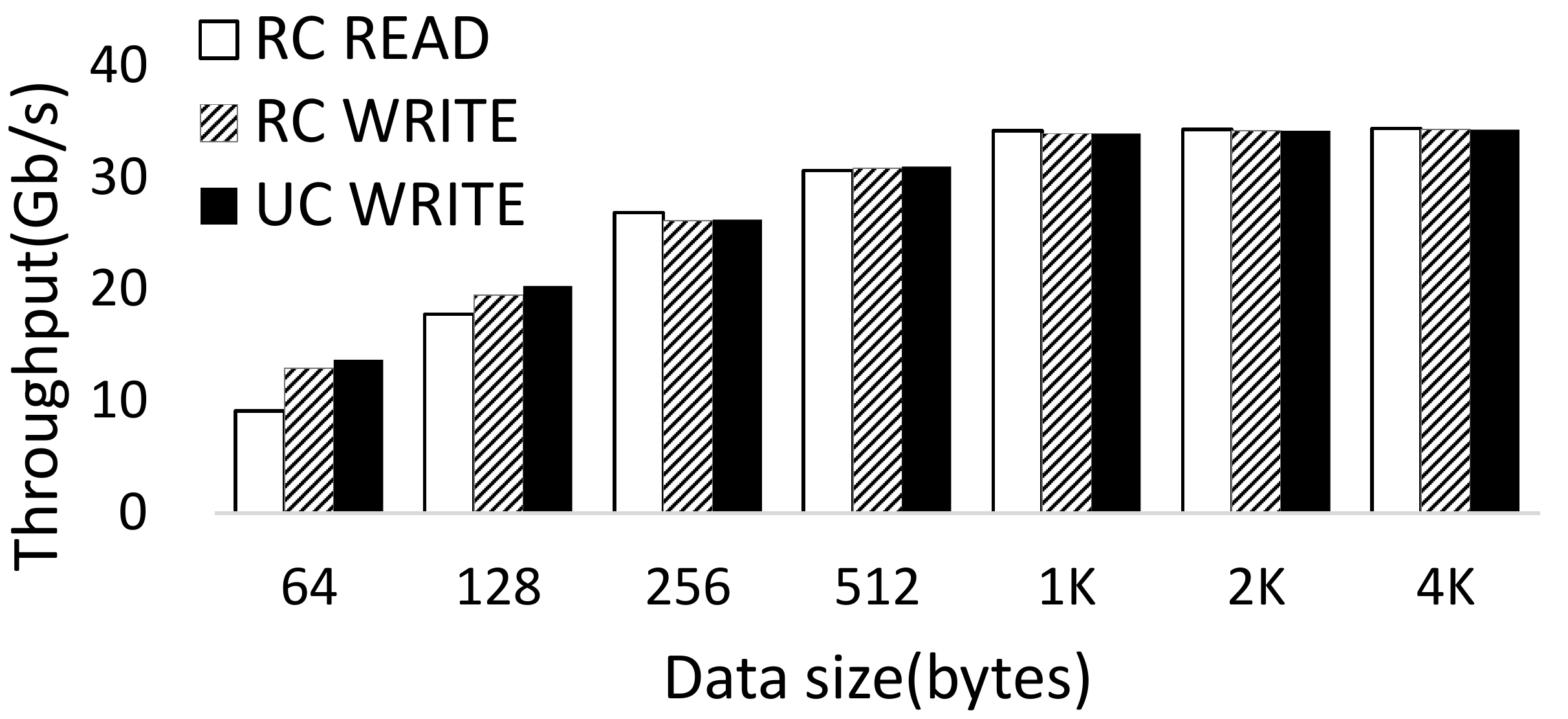}     
	}     
	\caption{Comparision of RDMA operations} 
	\label{fig:opts}     
\end{figure}

One-side verbs (Read and Write) bypass the remote CPU to operate directly on remote memory. This advanced feature is a popular option of data transfer in datacenters, because RDMA releases precious CPUs resource for the computation services. Our micro-benchmark shows that RC Write performs as well as UC Write. In Figure \ref{fig:opts}, the throughput of RC Read is close to RC write for large message. However, UC QPs does not support Shared Receive Queue (SRQ) \cite{specification}. Therefore we use RC as the default RDMA transport instead of UC when users decide to apply connected transport. The selection of RC Read and Write is adaptively adjusted based on the current CPU and memory consumption of servers (see Sec. \ref{sec:api}). To mitigate the impact of one-side verbs scalability, we introduce a lock-free approach for effective QPs sharing to support thousands of connections in one host. 

\subsection{Architecture and Programming Model}\label{sec:api}
\begin{figure}
	\centering
	\epsfig{file=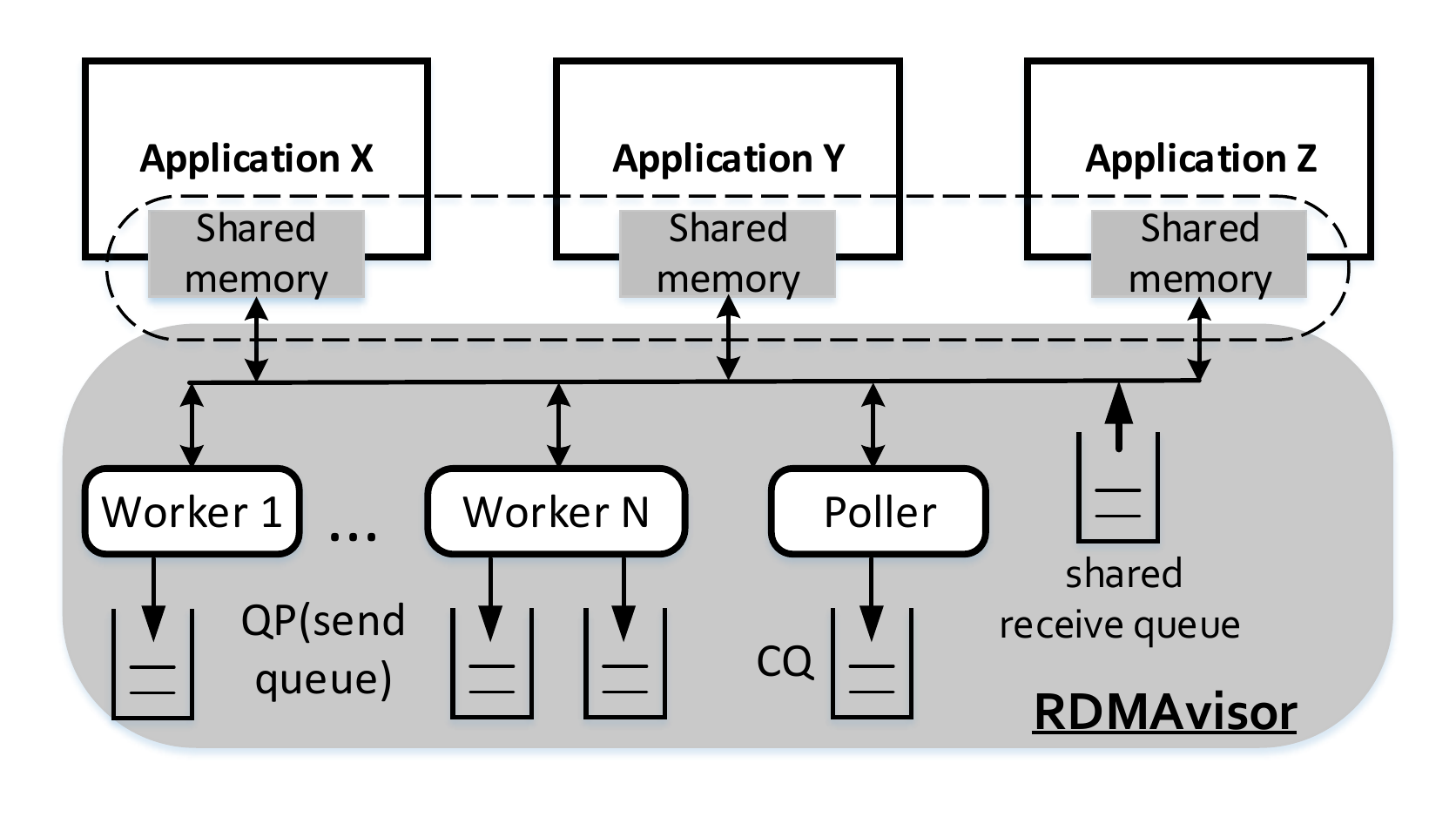, width=1.0\columnwidth}
	\caption{Overview of RDMAvisor}
	\label{fig:overview}
\end{figure}

Figure \ref{fig:overview} shows the prototype implementation of RaaS, \sysname\.  \sysname\ works as a daemon process to serve all applications through well designed interfaces. In \sysname, the working threads, named Worker, are responsible for  translating consumer's transmission request into appropriate WRs and then pushing WQEs  to the corresponding QPs. Another working thread, called Poller, is responsible for polling CQ and delivering message to the consumers in asynchronous mode. 

% managing QPs. 

% \textcolor{red}{To wangzhi : change this part }
% Generally, connected transports offer one-to-one communication between two queue pairs：$N*T$ QPs are required to connect $N$ machines and $T$ threads per machine, which may not fit in the NIC's queue pair chache. Threads can share QPs to reduce the QP memory footprint. QP sharing is typically implemented by creating several sets of $N$ QPs, where each set is connected to the $N$ machines \cite{d14farm}. Sharing QPs reduces CPU efficiency because threads contend for locks and the cache lines for QP buffers bounce between their CPU cores. 
% It works like system services, just calling the given interfaces, then all requests done by \sysname, in which the difference is the latter serving in mostly user space. Therefore, we believe it relived distributed system designers from seeking and designing high performance networking, which in our opinion should be a more general system infrastructure. 

\begin{figure}
	\centering
	\epsfig{file=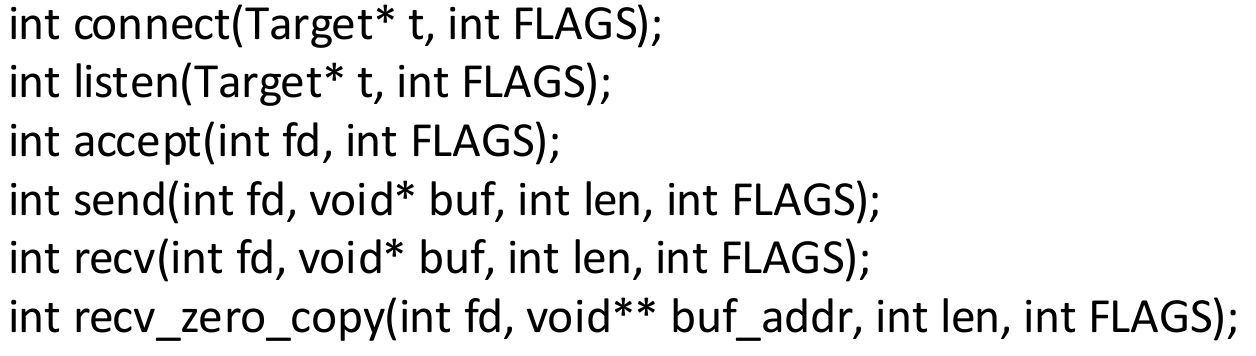, width=1.0\columnwidth}
	\caption{RaaS's API}
	\label{fig:api}
\end{figure}

% Normal users only need to select using reliable connection or unreliable datagram transport. 
\sysname\ supports simple interfaces to invoke RDMA functions by hiding the complicate details. Figure \ref{fig:api} shows the operations in {\sysname}'s interfaces. They are  socket-like operations which are friendly to users. Normal users can simplely use \texttt{send}, \texttt{recv} operations to transfer data. {\sysname} will adaptively select RDMA Send/Recv for data block of small size and RDMA Read/Write operations for large data according to the explicitly indicated data size in the parameter of `len'. Specifically, to reduce latency, \sysname\ chooses one-side verbs based on the current CPU consumption and work load, which is easily measured by the \sysname\ daemon process. 
It is notable that a parameter called \texttt{FLAGS} is introduced in the RaaS's interface. \texttt{FLGAS} is used to specific RDMA transport for one user with special requirement, e.g., $RC|WRITE$. This design provides the flexibility to help knowledgeable users to achieve customized setting.

The first three interfaces \texttt{ connection, listen, accept } are used to set up connections. Here \texttt{Addr} is an unified encapsulation of host address, including IPv4, IPv6 and RDMA's global\_id (GID) and local\_id (LID). The default value of FLAGS means RC operations in our design. Users can use \texttt{connect} to initiate active connections and \texttt{accept} to passively wait for new incoming connections. \texttt{send} is for outbound data transmission and users can use FLAGS \texttt{send} to explicitly specify the type of RDMA operation: SEND, RDMA READ or WRITE. Both \texttt{recv} and \texttt{recv\_zero\_copy} are used for receiving incoming data. The field of \texttt{fd} in \texttt{send}, \texttt{recv} and \texttt{recv\_zero\_copy} is used to specify on which logical connection the interface needs to be invoked. Notice that the interface of \texttt{recv\_zero\_copy} is provided for application which works in the blocking mode, i.e., it may take longer time to pick the data from the receive buffer of RDMA. By using \texttt{recv\_zero\_copy}, \sysname\ will delivers the incoming data directly from receiving buffers to the pre-registered private memory of the application, instead of occupying the shared memory of {\sysname}.

Notice that we present the \texttt{recv\_zero\_copy} interface, which is popular in high performance system. We don't provide \texttt{send\_zero\_copy} due to the revelation of related work \cite{f09min}. The reason is as following. For sending tasks, RNICs must get the physical address of outbound data to initiate a DMA operation. The memory region in which outbound data lying must be registered to RNICs beforehand. One way is to copy applications' outbound data to pre-registered memory region, called memcpy(). Another approach is to register the memory region of applications' outbound data to RNICs every time, called memreg(). However, in the case of relative small size of outbound data, memcpy() achieves better latency, but memreg outperforms memcpy when the size of outbound data increases. Therefore, \texttt{send\_zero\_copy} may be inefficient in sending relatively big size messages and \sysname\ is going to handle this according to related work \cite{f09min}.

\subsection{Scalability and Multiplexing of QPs}\label{sec:multiplex}
Due to the limitation of RC transport as aforementioned, we focus on solving the scalability problem of RDMA RC by introducing a more effective lock-free design of QPs sharing. The scaling problem incurs due to the limited NIC cache. If the accessing information is not cached in the NIC, it needs to fetch the desired data in memory and run a cache replacing algorithm. This process leads to the bottleneck and slows down the packets processing of RDMA transport. Many approaches have been proposed to reduce NIC cache misses, like using huge page to reduce mapping entries of physical address to virtual address \cite{d14farm}, adopting UD transport to reduce QPs needed by server applications \cite{k16fasst, k14keyvalue}. Our implementation also leverages the huge page and shared QPs. In what follows, we introduce the design of a lock free shared QPs approach.

As shown in Figure \ref{fig:overview}, for sending task, applications push requests to \sysname, and get responses from \sysname, both requests and responses are delivered by shared memory. This process is similar to the producer-consumer problem \cite{li2002analysis}. Applications produce send requests and \sysname\ consumes them. Since we intend to run RDMA operations in user space, \sysname\ applies shared memory with event descriptor for notification. Such operation reduces the involvement of kernel.  Therefore, we can significantly reduce the intra-process communication cost. Shared memory needs careful design of process synchronization. We combines it tactfully with event $fd$ to solve the producer-consumer problem mentioned before. Applications write send-requests to shared memory region, use event $fd$ to notify \sysname\, and then reads the same event $fd$ to get send result. For receiving task, Poller is responsible for polling CQ and notifies the corresponding application of inbound data available when getting CQEs. Therefore, our implementation processes data transmission tasks in a lock free manner and make them mostly done in user space. 

\begin{figure}
	\centering
	\epsfig{file=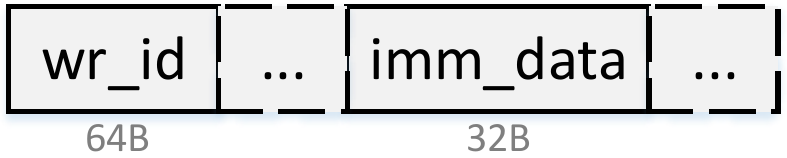, width=0.6\columnwidth}
	\caption{Data fields of work request}
	\label{fig:wr}
\end{figure}

Here, we explain how to identify the connection in the ``producer-consumer'' model.
In \sysname, all connections targeting at the same node are sharing one QP. We use the 4 byte data field in WQE to identify connections, called vQPN (virtual QPN). First, a vQPN is assigned to a new connection when it is generated. Then, \sysname\ places connections' vQPN in some fields of WQE when pushing WRs to queues. For one side verbs like RDMA Read and Write, \sysname\ can place connections' vQPN in the data field called \texttt{wr\_id} of WRs \cite{specification}, as shown in figure \ref{fig:wr}. After network interface card finishes provessing WRs from the send queues and generates relevant CQEs, vQPN can be accessed in \texttt{wr\_id} data field of the corresponding CQEs. By using the one side verbs, we do not need to keep the destination QP aware of anything, it is sufficient to achieve vQPN from \texttt{wr\_id}. However, for two side verbs, the destination needs to identify the source connection which is sharing by the source QP. In this case, we place vQPN in the data field called \texttt{imm\_data} in WRs \cite{specification}. Poller in the destination node is able to identify the source by accessing the vQPN of each connection in this field of CQEs, and then deliver data to the right logical application. As mentioned above, we place logical connections' id in WRs to recognize the connection in shared QPs. This operation is lock free and easy to implement. Besides reducing NICs cache misses, sharing QP can promote the probability of batching WRs, which in turn promotes throughput.

% zero overhead

% Related work also solve this problem by QP sharing through lock, which can be efficient if setting relevant parameters correct. These parameters include the size of cluster and how many threads share one QP. However, we provide a more simple and efficient way to share QP.

% \subsection{IPC and lock free concurrency control}\label{sec:ipc}

% work made design decisions based on the convincing picture of CPUs of server are busier than that of all clients, which is fixed, unsuitable in cloud computing because of resources utilization is time variant, making it unrealistic. It could be actually done in a more fine-grained way if system designers can be aware of the whole condition of the node even the cluster other than one application.

\begin{figure*}[ht] 
	\begin{minipage}[ht]{0.25\linewidth}
		\centering
		\includegraphics[width=.9\linewidth]{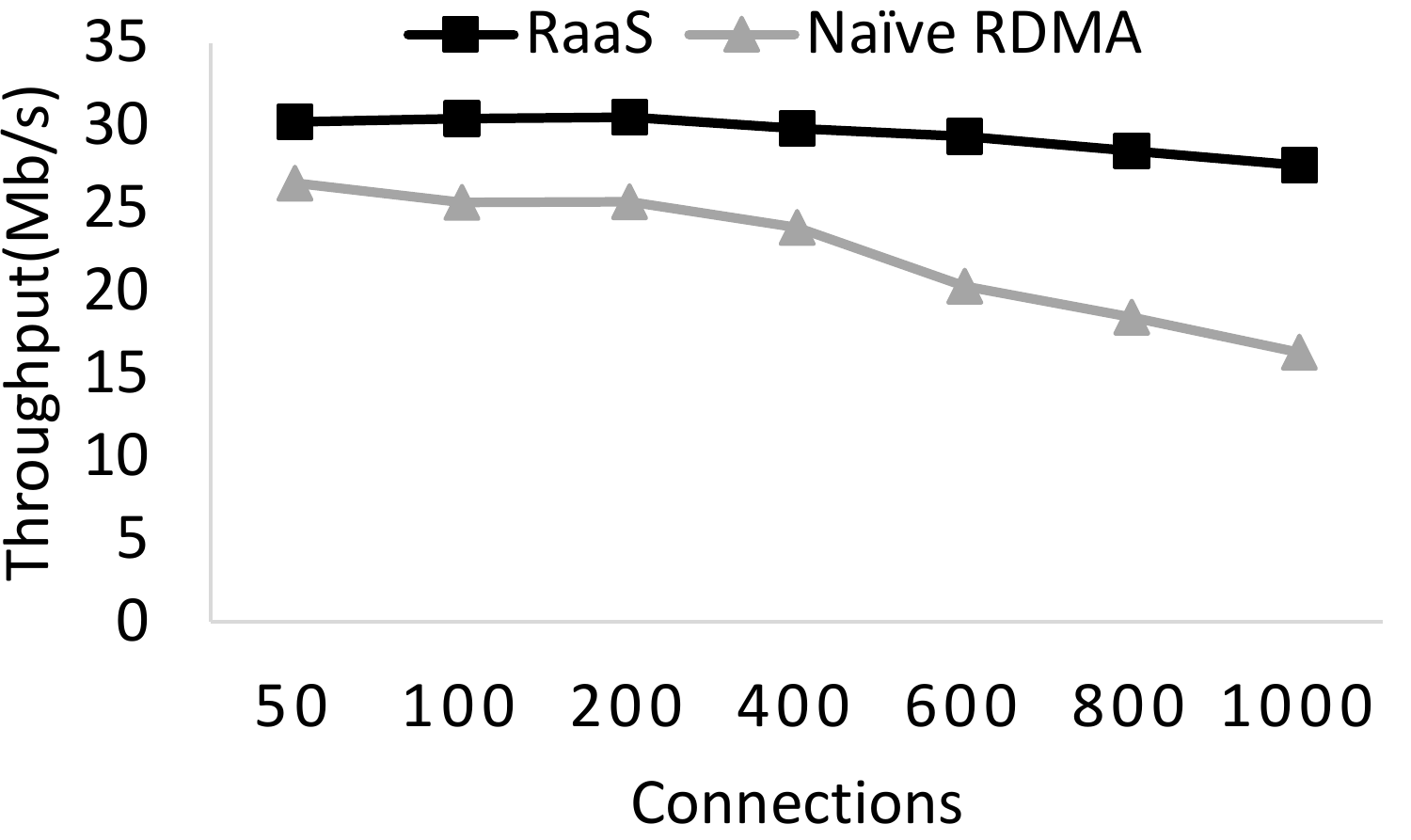} 
		\caption{Scalability}
		\label{fig:scalability}
		\hspace{0.3cm}
	\end{minipage}%%
	\begin{minipage}[ht]{0.25\linewidth}
		\centering
		\includegraphics[width=.9\linewidth]{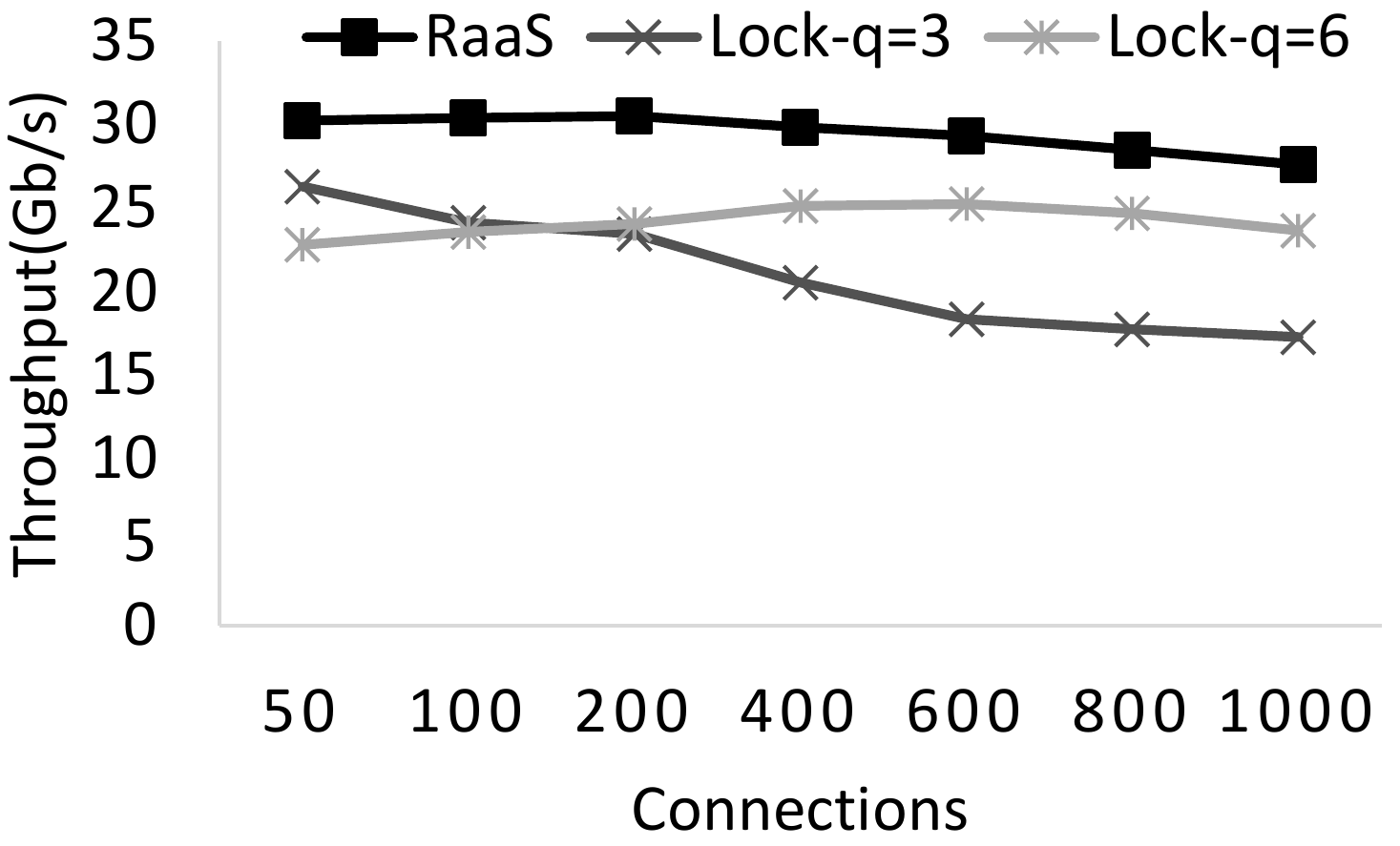} 
		\caption{Throughput vs. QPs sharing}
		\label{fig:lockfree} 
		\hspace{0.3cm}
	\end{minipage} 
	\begin{minipage}[ht]{0.25\linewidth}
		\centering
		\includegraphics[width=.9\linewidth]{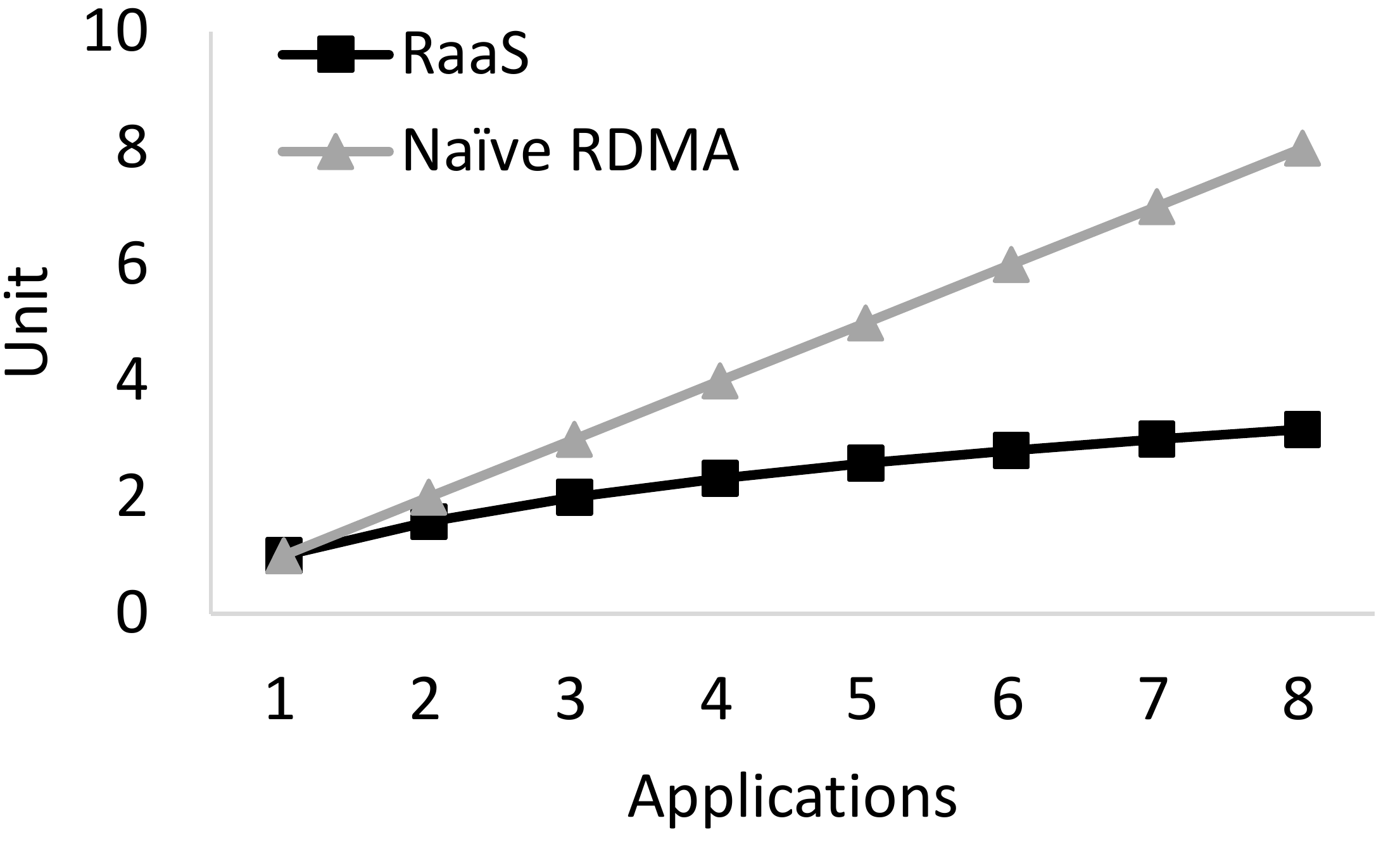} 
		\caption{Memory usage}
		\label{fig:memory}
		\hspace{0.3cm}
	\end{minipage}%% 
	\begin{minipage}[ht]{0.25\linewidth}
		\centering
		\includegraphics[width=.9\linewidth]{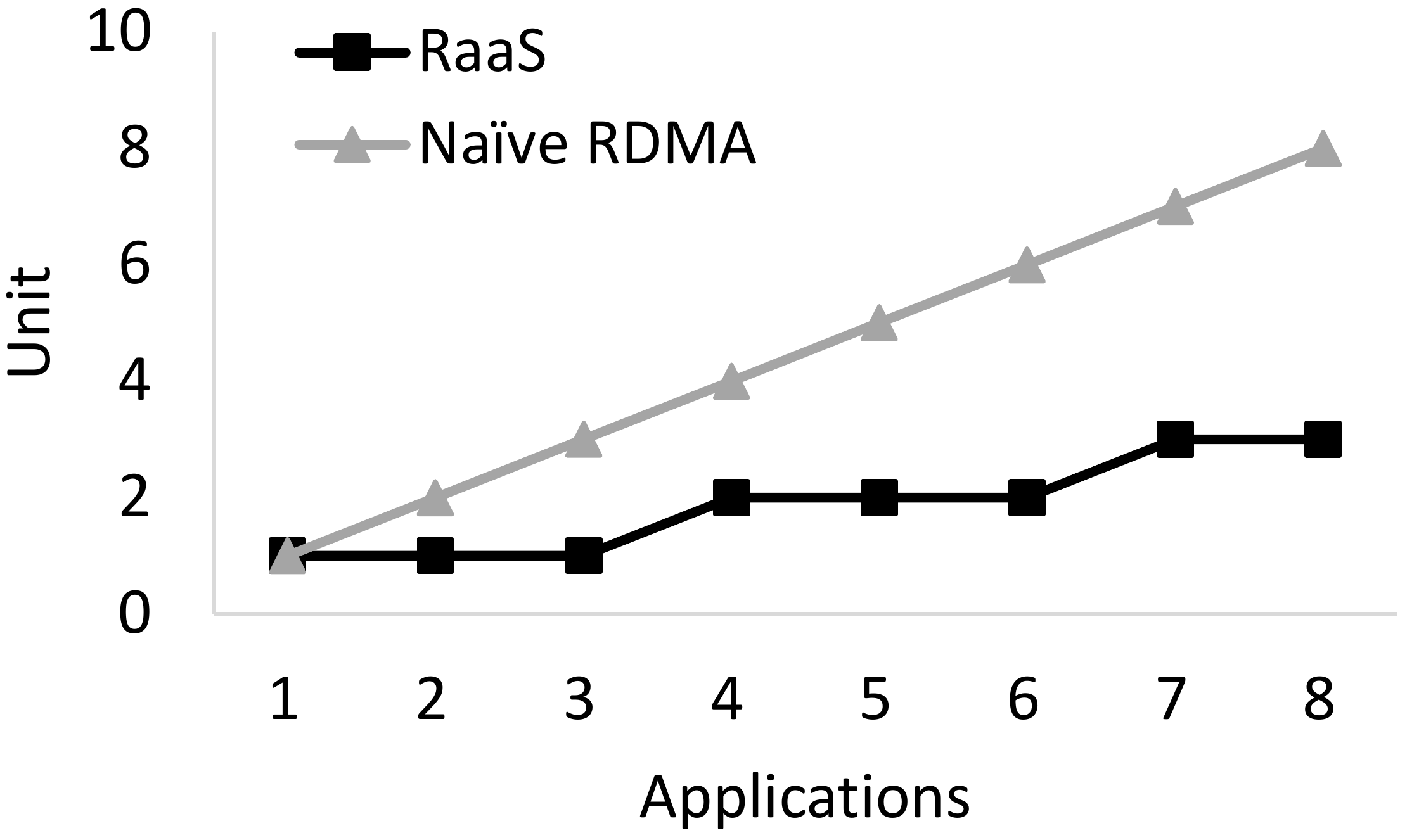} 
		\caption{CPU consumption}
		\label{fig:cpu}
		\hspace{0.3cm}
	\end{minipage} 
\end{figure*}

%\subsection{Resources sharing}
% Here, we explain the shared SRQs and CQs approach to improve the resource utilization. Resources are referred to memory and CPU here mostly. As we mentioned before, without \sysname, deploying RDMA enabled applications faces severe waste of resources because each application possess their own RDMA module, which can be sharing like receiving buffers, polling thread and so on. \sysname\ manages and controls resource sharing by combining basic assignment and demand assignment. Basic assignment means initial resources assignment if a new application requesting for RDMA service. Demand assignment is referred to the case that \sysname\ appends more resources when detecting there is performance dropping due to lack of some kinds of resources.

%% file: eval.tex
%!TEX root=main.tex

\section{Evaluation}

We evaluate the performance of {\sysname} on a cluster with four nodes. Each node ran CentOS 7.1 operating system on four 2.1 GHz Intel Xeon processors with a total of 24 cores. Each machine has 64 GB of memory and a 40 Gb ConnectX-3 RoCE network interface card. 

To evaluate the scalability of \sysname\ design, up to 1000 connections are generated on one machine to randomly read 64KB data from other machines. We compare the RaaS Read operations with the naive RDMA Read verbs where the QPs are not shared by connections. As shown in Figure \ref{fig:scalability}, the throughput of naive RDMA starts to drop when the size of connections, i.e., QPs, exceeds 400. RaaS shows stable performance since \sysname\ allow reuse of QPs to communication with multiple other QPs. It is notable that \sysname\ also outperforms the naive RDMA when the number of connections is even less than 400. This is because the reuse of QPs has higher opportunity of batching WRs than the case of one QP per thread.

We then compare RaaS with the design of sharing QPs with locks \cite{d14farm}, where each QP is shard by $q$ threads. Figure \ref{fig:lockfree} shows results when $q$ is 3 and 6 by running random read. We can see that sharing QPs can reduce throughput when threads contend for locks. RaaS can mitigate the impact of lock contention. Specifically, our design is not sensitive to the number of $q$ and the scale of clusters \cite{d14farm}.  

We then compare the RaaS resource allocation with the applications using naive RDMA. The resource (memory and CPU) consumption is normalized. Here, one unit is referred to the resource required by one application to set up the connection. Figure \ref{fig:memory} shows the memory consumption and Figure \ref{fig:cpu} shows the CPUs consumption. We can see clearly that the resource required by the naive RDMA operations increases linearly with the increase of applications. However, the design of RaaS propose more effective usage of resource which results in a slow growth of memory/CPU. With the increase of connections, RaaS has a higher probability to realize QPs sharing.

%% file: conclude.tex
%!TEX root=main.tex
\section{Conclusion}
In this paper we presented a simple and scalable RDMA as Service (RaaS) to mitigate the impact of RDMA operational details. RaaS provides careful message buffer management to improve CPU/memory utilization and improve the scalability of RDMA operations. These optimized designs lead to simple and flexible programming model for common and knowledgeable users. We have implemented a prototype of RaaS, named {\sysname}, and evaluated its performance on a cluster with a large number of connections. Our experiment results demonstrate that {\sysname} achieves high throughput for thousand of connections and maintains low CPU and memory overhead through adaptive RDMA transport selection. Part of this work has been devoted to the Ceph community and the source code is available at https://github.com/wangzhiNJU/ceph/tree/rdma. 